\documentclass[aps,twocolumn,showpacs,amsmath,amstex,amssymb,citeautoscript]{revtex4-2}
\usepackage{physics}
\usepackage[colorlinks,linkcolor=magenta,citecolor=magenta]{hyperref}
\usepackage{bm}

\usepackage{graphicx}
\usepackage{dcolumn}
\usepackage{bm}
\usepackage{dsfont}
\usepackage{amsthm,amsmath,amssymb}
\usepackage{exscale}
\usepackage{relsize}
\usepackage{xcolor}
\usepackage{appendix}
\usepackage{mathrsfs}

\usepackage{tikz}

\usepackage{lipsum}

\begin{document}

\title{Ferromagnetic Traps for Quasi-Continuous Operation of Optical Nanofiber Interfaces}
\author{Ruijuan Liu$^{1,3}$}
\email{rjliu20@fudan.edu.cn}
\author{Jinggu Wu$^{1,3}$}
\author{Yuan Jiang$^{2,4}$}
\author{Yanting Zhao$^{2,4}$}
\email{zhaoyt@sxu.edu.cn}
\author{Saijun Wu$^{1,3}$}
\email{saijunwu@fudan.edu.cn}

\affiliation{$^1$Department of Physics, State Key Laboratory of Surface Physics and Key Laboratory of Micro and Nano Photonic Structures (Ministry of Education), Fudan University, Shanghai 200433, China.
}
\affiliation{$^2$State Key Laboratory of Quantum Optics Technologies and Devices, Institute of Laser Spectroscopy, Shanxi University,Taiyuan 030006, China.}
\affiliation{$^3$Shanghai Key Laboratory of Metasurfaces for Light Manipulation, Shanghai 200433, China.}
\affiliation{$^4$Collaborative Innovation Center of Extreme Optics, Shanxi University, Taiyuan 030006, China}
\date{\today}

\begin{abstract}

A soft ferromagnetic plate uniformizes Tesla-level fields generated by attached permanent magnets, producing a smooth and electronically tunable surface field on the opposite side. By arranging $n$ precisely fabricated rectangular plates, a nearly ideal magnetic quadrupole field with a substantial gradient can be created at center. This robust and rapidly tunable field configuration is well suited for two-dimensional magneto-optical trapping (2D-MOT) and magnetic guiding of cold atoms. By aligning an optical nanofiber (ONF) along the zero-field line of a planar 2D-MOT in a 2-plate assembly, we demonstrate quasi-continuous, field-free operation of the quantum optical interface without switching off the magnetic field. Transient transmission spectroscopy with nanosecond laser pulses is performed on the $^{87}$Rb D2 line at a measurement repetition rate as high as 250~kHz. The observed line broadening, while not yet fully understood, is partially attributed to residual magnetic fields in the $n=2$ assembly. Through additional measurements and simulations, we verify that these residual fields can be fully eliminated in an $n=4$ assembly, resulting in an ultra-straight 2D trap that supports uniform light–atom interaction over exceptionally long, field-free distances. We extend our discussion to $n=6$, $n=8$ designs with similar uniformity but multiple zero-field lines. With its strong gradient for magnetic trapping, the ferromagnetic devices also enable new quantum optical scenarios featuring interactions between co-guided atoms and photons at ONF interfaces. 
\end{abstract}
\maketitle


\section{Introduction}

Optical nanofibers (ONFs) have become integral components in the field of quantum optics~\cite{hakuta_manipulating_2012,Solano2017,Solano2017b,Sheremet2023,Zhang2024,Jain2024}, primarily due to their ability to support highly efficient interactions between guided photons and surrounding near-field atoms. These interactions are sustained over exceptionally long distances, denoted as $l$, which are critical for realizing strong, collective light-atom coupling. However, achieving such precise interactions necessitates maintaining the ONF interaction zone free from any uncontrolled level-shifting fields, which can otherwise distort the atomic energy levels and broaden the coupling strengths. Additionally, keeping the ONF interface field-free is important for precision atomic spectroscopy at the nanoscale~\cite{Morrissey2013, Solano2019c},  and for sub-Doppler cooling~\cite{MetcalfBook} which is an essential step to load a lattice~\cite{Vetsch2010,Goban2012a,Meng2018,Su2019,Kestler2023}. However, laser cooling starts with magneto-optical trapping (MOT)~\cite{MetcalfBook}. For efficient spatial confinement of the atomic sample, MOT operation requires magnetic field gradients on the order of 10~G/cm. This leads to Gauss-level average fields along the ONF interface even for $l$ of merely a few millimeters. Switching off the MOT field to prevent such perturbations can be too slow in many experimental setups~\cite{Sague2007,Corzo2019,Vetsch2010,Solano2019c,Solano2017,Patterson2018}, particularly since these measurements are often at single photon level and requires many repetitions for enough statistics.

\begin{figure*}[htbp]
        \centering
        \includegraphics [width=1\linewidth]{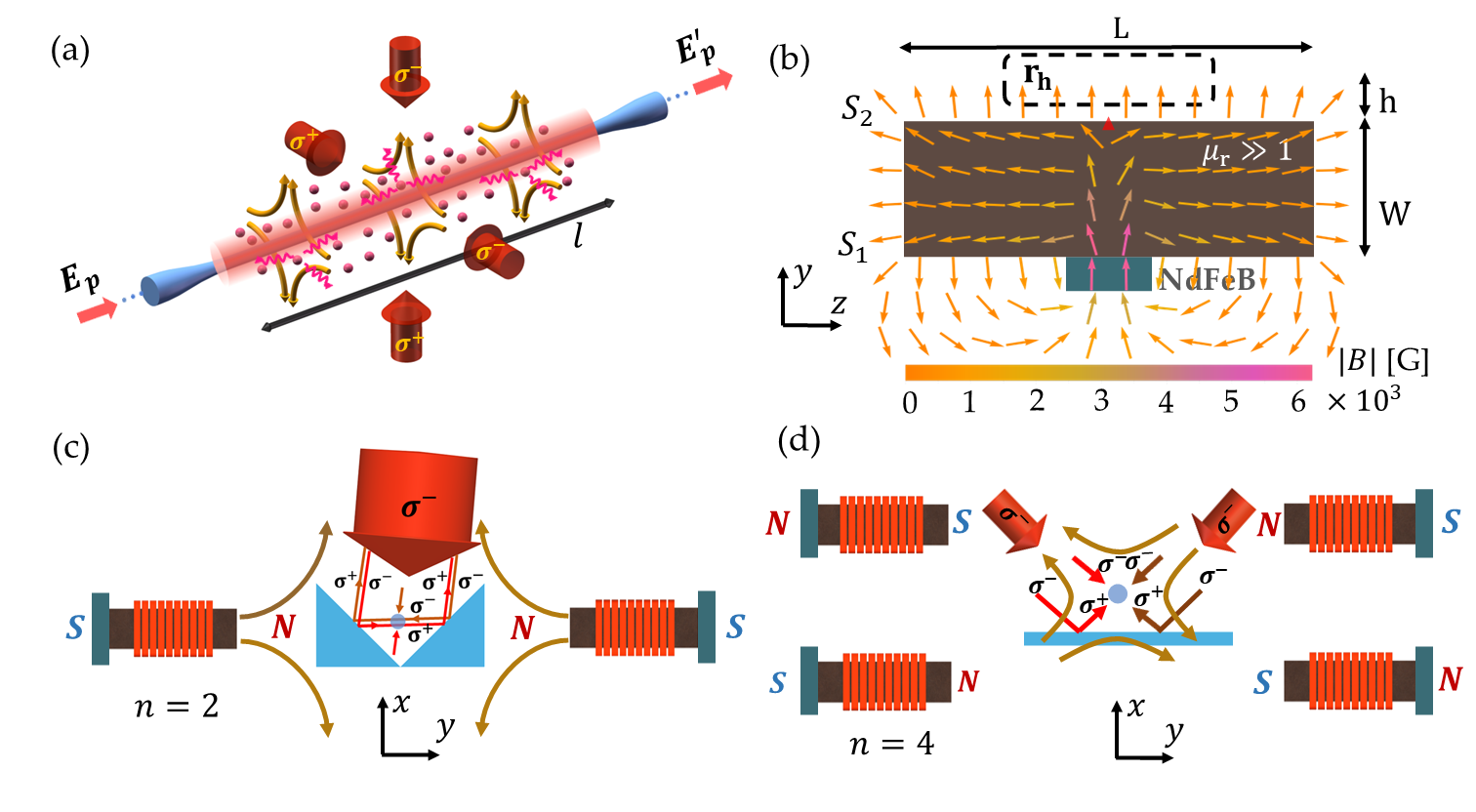}
        \caption{(a): Schematic of an ONF-2D-MOT interface with interaction length $l$. (b): Uniformization of magnetic field by a rectangular ${\bf \mu}-$metal plate. For the precisely machined plate, the magnetic field within the dashed line box on top can be highly uniform, being nearly invariant along $z$.
        The center of the $S_2$ surface ${\bf r}_s$ is marked with a red triangle marker.
        (c): The prism 2D-MOT using the $n$=2 plate structure in this work. Local laser beam polarizations defined along $+x,+y$ axes are marked. (d): 2D-MOT using $n$=4 plate structure. Local laser beam helicity are defined along $+x,+y$ axes in (a,c), and along a pair of $45^{\circ}$ axes in (d). In (c,d) the blue disk in the MOT center represents ONF. Its sub-micron diameter is exaggerated for clarity.}
       \label{figSetup}   
    \end{figure*}

A potential solution to meet the field-free requirements without switching off the field is to load the ONF interface with a two-dimensional MOT~\cite{vengalattore_enhancement_2004}. As in Fig.~\ref{figSetup}a, 2D-MOT operates with magneto-optical trapping force in the $x-y$ plane only, without requiring any field gradient along $z$. Cold atoms trapped in 2D-MOT is highly elongated, naturally serve to load the ONF interface. By precisely locating the ONF to the zero-field line, the magnetic field sensed by cold atoms in the nanoscale near field is practically zeroed, even for 2D-MOT operates with very high field gradients~\cite{wu_demonstration_2007}. Importantly, since polarization gradient cooling is most efficient at zero field~\cite{Townsend1995, vengalattore_enhancement_2004}, we expect efficient loading of atoms to the ONF optical lattice~\cite{Vetsch2010,Meng2018,Su2019,Kestler2023}. However, all these 2D-MOT benefits require precise alignment of the zero-field line with the long ONF. While ONF is itself highly stiff~\cite{Zhang2024} to support a straight line within {\it e.g.} a centimeter, a 2D-MOT optimized for free-space quantum optics~\cite{Zhang2012a}  does not guarantee a straight enough zero-field line to match the straight ONF. In addition, to construct a 2D-MOT usually requires closely spaced rectangular coils carrying high currents~\cite{Zhang2012a}, a setup that may be challenging to implement to ONF-based quantum optical platforms~\cite{Russell2013,russell_sub-doppler_2012,Kumar2015} where the ONF mounting structures  necessarily constrains the size and complexity of the setup.

In this work, we draw inspiration from earlier developments in magnetically guided atom interferometry~\cite{wu_demonstration_2007} and use ferromagnetic trap for field-free operation of ONF interfaces. As in Fig.~\ref{figSetup}(b), soft-ferromagnetic ${\bf \mu}-$metal plate poled by permanent magnets from one side can generate highly uniform surface field on the other~\cite{JacksonBook,coey2009Book}. By properly arranging $n=2$ (Fig.~\ref{figSetup}c) or $n=4$ (Fig.~\ref{figSetup}d) plates, a zero-field line is formed near the symmetry center to support the 2D-MOT operation. This zero-field line is ultra-straight~\cite{wu_demonstration_2007}. Its position can be finely adjusted via the electronic current surrounding the plates. By shifting this zero-field line to overlap with the straight ONF, the magnetic field is nullified microscopically to support field-free operation without switching off the magnetic field for the 2D-MOT. We experimentally demonstrate the key ingredients of our proposal with a compact two-plate ONF interface, featuring microsecond-level cooling-measurement cycle time. An anomalous line-broadening in transient ONF spectroscopy is observed, which is not fully understood (Appendix~\ref{sec:lwApp}), but is partly attributed to a residual field gradient along the zero-field line direction $z$ (Fig.~\ref{figField}a) in the $n=2$ assembly. This $z$-gradient can be completely eliminated in the $n=4$ configuration (Fig.~\ref{figField}(c,d)), with which we discuss on the prospects for quasi-continuous, field-free operation of ONF-cold-atom interface with $l\geq$100~mm interaction distances, and to enable novel quantum optical research featuring strongly coupled photons with matterwave in their respective waveguides. 

In the following our work is presented in three sections. First, in Sec.~\ref{sec:nplate}, we outline the general principles of constructing an ultra-straight ferromagnetic trap for 2D cooling and trapping and for integrating the device with an optical nanofiber. In Sec.~\ref{sec:exp0}, we present our experimental demonstration on field-free operation of the ONF interface, in particular, on a quasi-continuous transient absorption spectroscopy measurement with a repetition rate as high as 250~kHz. In Sec.~\ref{sec:straight} we clarify practical aspects of the ultra-straight 2D field generation and propose an ONF interface with up to $l=10~$cm field-free interaction length. We summarize this work in Sec.~\ref{sec:sum} with an outlook into prospects of combining the ONF-technology with the ferromagnetically generated, nearly ideal 2D trapping potentials, particularly for interfacing co-guided photons and atoms. 

\section{Ferromagnetic assembly for 2D cooling and trapping}\label{sec:nplate}    

\subsection{Robust uniformization of magnetic field}\label{sec:uniform}

The ${\bf \mu}-$metal is a nickel-iron-based soft-ferromagnetic alloy featuring high permeability ${\bf \mu}={\bf \mu}_r {\bf \mu}_0$, high saturation field $B_{\rm sat}$, and low remanence field $B_{\rm rem}$. The vacuum permeability is ${\bf \mu}_0=4\pi\times 10^{-7}~{\rm N/A^2}$. For example, the ${\bf \mu}-$metal material considered in this work is composed of 80\% Nickel and 15\% Iron. It has a saturation field of $B_{\rm sat}\approx 0.8$~T, a permeability of ${\bf \mu}_r\sim 10^5$ at $B\ll B_{\rm sat}$, and $B_{\rm rem}\sim$ 1~Gauss~\cite{coey2009Book}. 

\begin{figure}[htbp]
        \centering
        \includegraphics [width=1\linewidth]{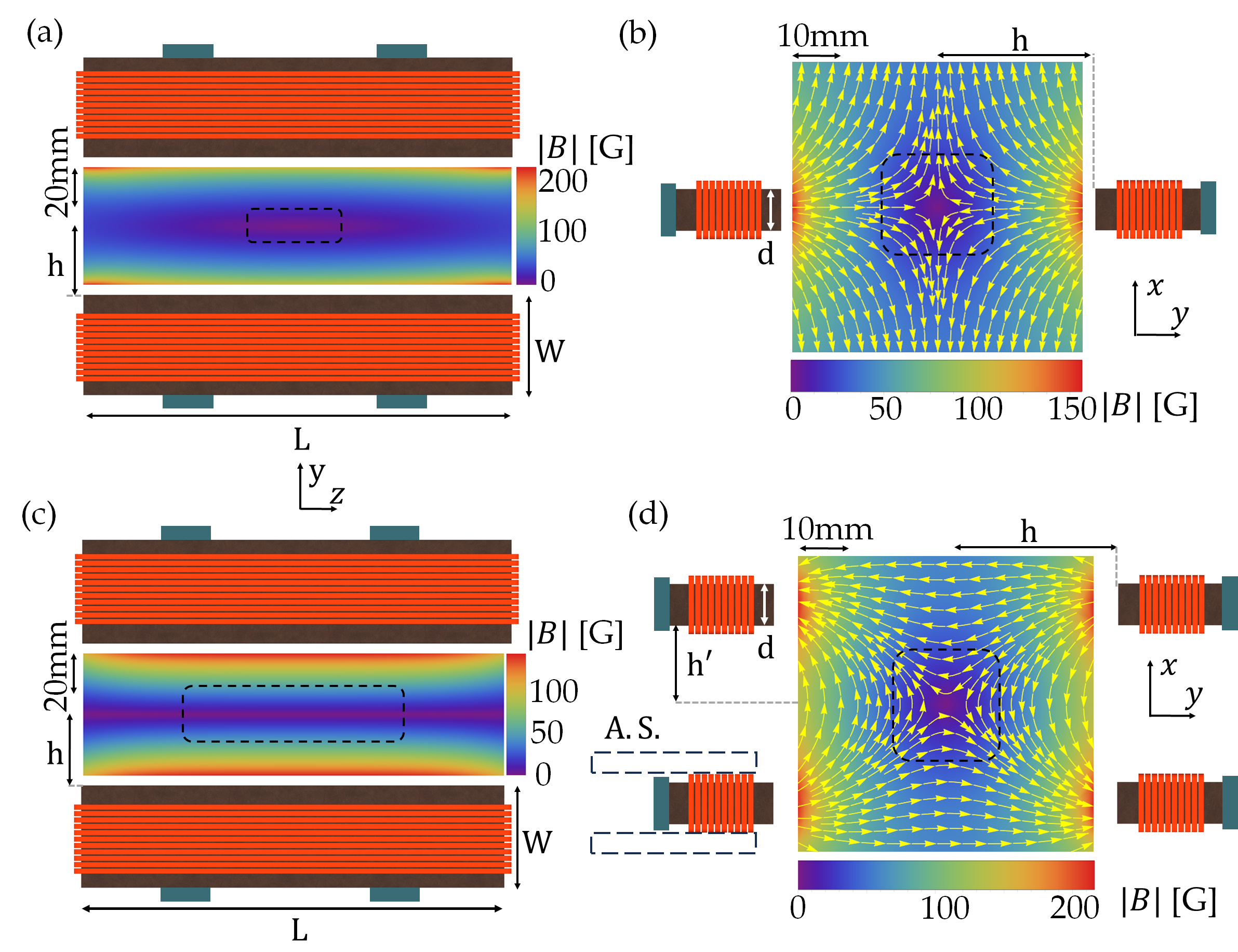}
        \caption{Numerical simulation of the 2D-quadruple field~\cite{Elleaume1997,Chubar1998} for the 2-plate assembly, (a,b), according to Sec.~\ref{sec:exp}; and 4-plate assembly, (c,d), according to Sec.~\ref{sec:perm}.  Here $L=120~$mm, $W=40~$mm, $d=10~$mm. See Appendix~\ref{sec:simu} for simulation details.  The plate dimensions are not to scale. A.S.: Optional, additional shielding of source magnet variations for operating 2D-trap at large $h,h'$ distances~\cite{foot:closure}. }
         \label{figField} 
    \end{figure}

As in Fig.~\ref{figSetup}b, we consider rectangular ${\bf \mu}-$metal plate with width $W$, length $L$, and thickness $d\ll W, L$. The plate is magnetized primarily by permanent magnets, {\it e.g.} a Neodymium Iron Boron (NdFeB) block, from the $S_1$ side. To provide electronic tunability, the plate is in addition poled by current-carrying coils surrounding the plate (see e.g. Fig.~\ref{figField}a). The NdFeB magnet with size $a\approx d$ is small. In absence of the ${\bf \mu}-$metal, its field at distance $W$ away is effectively a dipole field which varies substantially over any $O(W)$ distance. However, as highlighted by the dash-lined box in Fig.~\ref{figSetup}b, the field flux after being guided through the ${\bf \mu}-$metal  becomes uniform near the distant surface $S_2$. Here, ${\bf r}_h={\bf r}_s+(x_h,y_h,z_h)$ with $|x_h|,|y_h|=O(h)\ll W$ from the plate surface, and $|z_h|<L/4$ to avoid edge effects. 


In Sec.~\ref{sec:perm}, we present a detailed study of the uniform magnetic field generated by the NdFeB$-{\bf \mu}$-metal plate structure. In particular, the uniform field distribution near the $S_2$ is robust, insensitive to variations in magnetization near the $S_1$ surface, whether induced by permanent magnets or current-carrying wires, as well as by spatial variations in the permeability ${\bf \mu}_r$ of the ${\bf \mu}-$metal itself. The field distribution can therefore be replicated by multiple plates in the $n=2$ (Fig.~\ref{figSetup}c) or $n=4$ (Fig.~\ref{figSetup}d) configurations to generate the central zero-field line for the 2D-MOT operation.

Here, we provide an intuitive explanation on the robust field uniformity, which is analogous to field uniformization by an isolated conductor in electrostatics~\cite{JacksonBook}.
Specifically, the ${\bf \mu}-$metal plate responses to the source field with a redistribution of effective magnetic charge. Similar to electrostatic shielding by a grounded conductor, the magnetic charge redistribution shields the inhomogeneous source field near $S_1$ from affecting the proximity of $S_2$. The shielding is therefore highly efficient for plate with large enough $W$, $L$, and $d$. Certainly, to keep the magnetic charge neutrality, there are extra charges in the ${\bf \mu}-$metal, which tend to distribute uniformly on the surface to minimize the field energy~\cite{JacksonBook}. This surface charge picture becomes accurate near $S_2$ where $B \ll B_{\text{sat}}$ so that the ${\bf \mu}-$metal response is linear. In addition, with ${\bf \mu}_r\gg1$ in the ${\bf \mu}-$metal, the ${\bf B}$ field in air (or vacuum) is perpendicular to the surface, according to the tangential ${\bf H}=\frac{1}{{\bf \mu}}{\bf B}$ continuity. As in electrostatics, the uniform surface charge leads to highly uniform ${\bf B}$ near the flat $S_2$ surface, insensitive to small variations of source fields or the ${\bf \mu}_r$ distribution. The uniformization is also effective against imperfect current-coil winding~\cite{wu_demonstration_2007}. This effect is most pronounced for thin-wire winding at a sufficient distance from the $S_2$ surface, as illustrated in Fig.~\ref{figField}(c,d).  While the shielding of source magnetic irregularities tends to degrade at large working distances, the shielding can be reinforced, if necessary, by constructing additional, ${\bf \mu}-$metal enclosures for the sources (Fig.~\ref{figField}d).

Finally, the field distribution is sensitive to the surface shape of the ${\bf \mu}-$metal itself, which can be machined with micrometer precision, over large scale. Mesoscopically, surface corrugations with a characteristic length scale $\xi$ of micrometers are typically present in these plates, but the associated fringe fields decay exponentially along $y$~\cite{Wang2004, ThesisWu}. Therefore, to mitigate the effects of fringe field irregularities, 2D trapping should be positioned at a distance $y_h = h \gg \xi$.


\subsection{$n=2$ plate trap}
As in Fig.~\ref{figSetup}c, Fig.~\ref{figField}(a,b) and Fig.~\ref{figMOT}, we consider arranging $n=2$ plates symmetrically with respect to the $y=0$ plane with a gap distance of $\Delta y=2h$. With the opposite magnetization, the fields generated by the two plates cancel at ${\bf r}_0=(0,0,0)$ to form a magnetic zero. Near the center, the field ${\bf B}=B_1^x x {\bf e}_x+B_1^y y {\bf e}_y+B_1^z z {\bf e}_z$, with gradients $B_1^x:B_1^y:B_1^z=(1-\varepsilon):-1:\varepsilon$. Here $\varepsilon\sim (h/L)^2$ when $\varepsilon\ll1$. The field zero is therefore a three-dimensional (3D) one. The residual $B_1^z$ gradient is associated with the fact that the field contributions from the two plates add up along $z$ for off-center locations. Nevertheless, with an operation distance $\varepsilon=h/L\ll 1$, the field distribution becomes highly elongated along $z$ to mimic a 2D field, as suggested by the dashed-line box in Fig.~\ref{figField}(c,d), with an approximate zero-field line near the center.

\subsection{$n=4$ plate trap}
As in Fig.~\ref{figSetup}d and Fig.~\ref{figField}(c,d), we now consider arranging $n=4$ plates symmetrically in the $x-y$ plane, with a gap distance of $\Delta y=2h$ and $\Delta x=2 h'$ (Fig.~\ref{figField}b). As in Fig.~\ref{figField}c, in contrast to the $n=2$ case, in the $n=4$ assembly the field variation along $z$ caused by each plate due to the finite $L$ are completely canceled at $x=y=0$. This results in a quadrupole field distribution, ${\bf B}=B_1 y {\bf e}_x+B_1 x {\bf e}_y$ at the center. Furthermore, with $h,h'\ll W$, the gradient  $B_1$ varies extremely slowly within {\it e.g.} $|z|<L/4$, as suggested by the dashed-line box in Fig.~\ref{figField}(c,d).

\subsection{Electronic tuning of field offset}
The magnetic response of $\mu$-metal is quite fast~\cite{coey2009Book}. Building on earlier ferromagnetic trapping techniques~\cite{vengalattore_enhancement_2004,wu_demonstration_2007}, we combine permanent magnetization with electronic tuning to generate strong and adjustable quadrupole field gradients $B_1$ at large distances $h$. As illustrated in Fig.~\ref{figSetup} and demonstrated by the data in Fig.~\ref{figMOT}f, small adjustments to the currents surrounding the plates can offset the magnetization of the $\mu-$metal and shift the zero-field line. By varying the currents in selected plates of the $n$-plate assembly in Fig.~\ref{figField}, the position of the zero-field line can be shifted over millimeter distances in the $x$-$y$ plane within tens of microseconds, with high precision and repeatability. By aligning the local zero-field line with the ONF position, field-free operation of atomic physics or quantum optics experiments can be performed at the ONF interface.  On the other hand, uniform in-plane field offsets within ten-Gauss level can be applied for {\it e.g.}, canceling local vector shifts naturally exits at nanophotonic interfaces~\cite{Zhou2024,Pache2025} or simply for atomic state preparation. 


Finally, while applying a uniform magnetic offset in the $x$-$y$ plane is feasible, achieving a uniform $B_z$ component in the $n$-plate geometry is more challenging. Even a spatially uniform external field can induce non-uniform magnetization responses in the $\mu$-metal, compromising the uniformity of the field along the $z$-axis. Nevertheless, for sufficiently large plate structures, the non-uniformity of the applied field scales with the ratio $l/L$,  across the ONF interaction length $l$, and can therefore be quite negligible for most transient applications.

\subsection{2D-MOT-based ONF interface}\label{sec:MOT}

\begin{figure*}[htbp]
        \centering
        \includegraphics [width=1\textwidth]{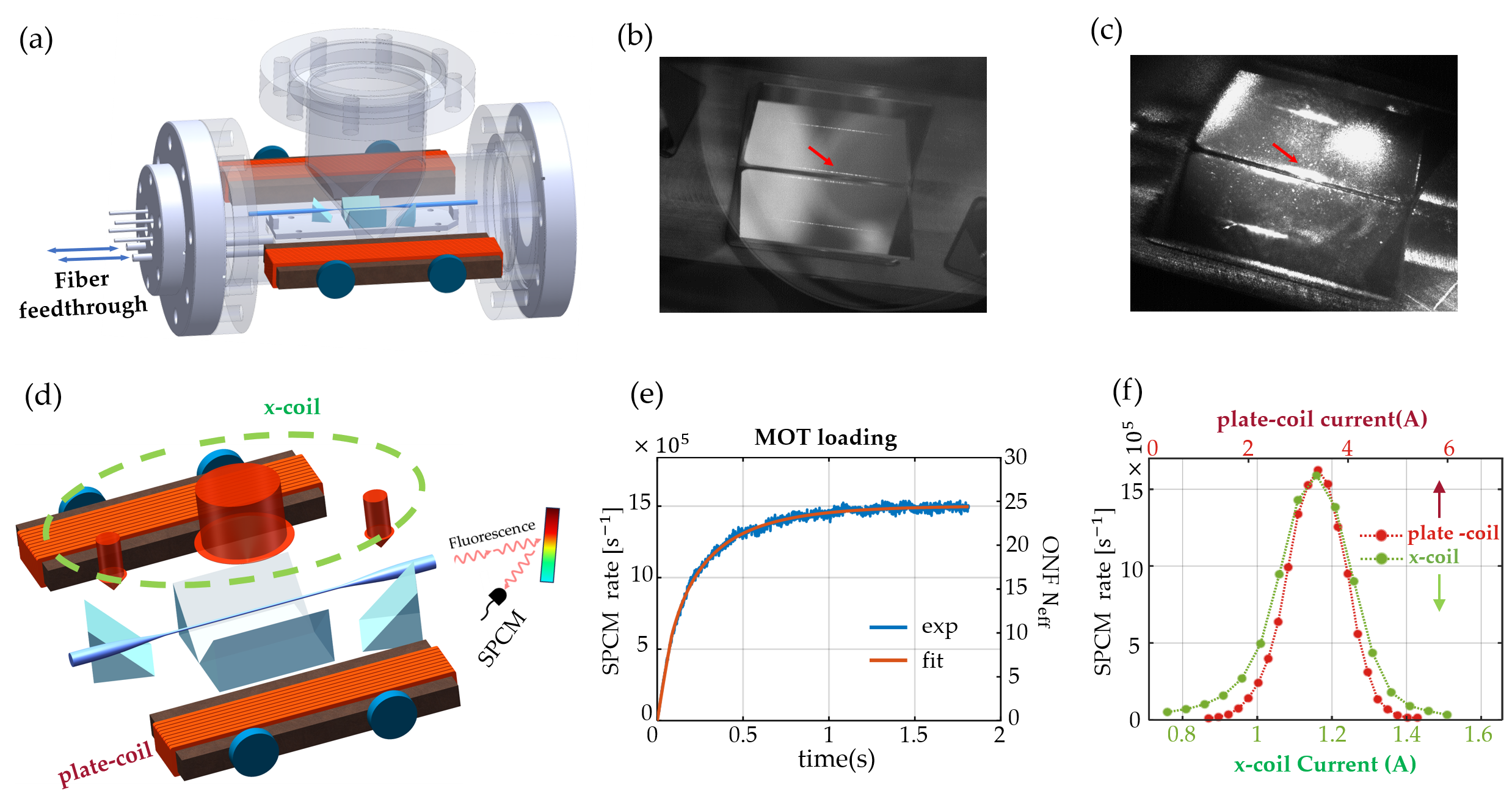}
     \caption{(a): Schematic of the ONF-2D-MOT apparatus in this work. (b): The ONF-2D-MOT interface through the vacuum window. The ONF is visible through Rayleigh scattering of a 100~$\rm {\bf \mu} W$ heating light at 843~nm. (c): The 2D-MOT (marked with red arrow) and its three mirror images. (d): Schematic of the ONF-2D-MOT in planar geometry aided by two pairs of right-angle prisms. The dash circle represents the $x-$coil with diameter $D=280~$mm which is 35~mm away from ONF. (e): Time-dependent fluorescence count through ONF and effective atom number $N_{\rm eff}$ during the MOT loading. The fit model for loading the density-limited MOT~\cite{Townsend1995} (Appendix~\ref{secFit}), with $\rho_0=3\times 10^{10}/{\rm cm}^3$ final density near ONF, suggests a MOT lifetime of $\kappa^{-1}=0.4~{\rm sec}$. (f): Steady-state fluorescence count as a function of 2D-MOT position, adjusted by the $x$-coil and the plate-coil currents.}
      \label{figMOT} 
 \end{figure*}
 
Both the quasi-2D field as in Fig.~\ref{figField}(a,b) and the 2D field as in Fig.~\ref{figField}(c,d) can be combined with properly polarized laser beams for magneto-optical trapping. 

For the two-plate assembly, the easiest way to form the quasi-2D-MOT is to use a right-angle prism pair that retro-reflect a single beam (Fig.~\ref{figSetup}c). The confinement along the $z-$direction is provided by two additional beams, also with two prisms in this work (Fig.~\ref{figMOT}d). This planer arrangement can be conveniently integrated with an ONF mount~\cite{Su2019} (Fig.~\ref{figMOT}(a-c)), as suggested by Fig.~\ref{figSetup}c. By adjusting the plate-coil current to shift the field along $y$, and an $x-$coil current to shift the field along $x$, the zero-field line is precisely adjusted in the $n=2$ assembly to overlap with ONF. Clearly, the two-plate configuration only supports a quasi-2D-MOT. Depending on the $\varepsilon=h/L$ ratio, the weak gradient along the ONF direction may limit the field-free ONF-interaction length $l$. 

On the other hand, as illustrated with Fig.~\ref{figField}(c,d), a true 2D quadruple field can be formed with the $n=4$ plate assembly. Similar to Fig.~\ref{figSetup}c, one prefers a planer geometry for the 2D-MOT construction, which can then be integrated with a standard ONF-mount~\cite{Su2019}. Here, with the field lines rotated by $45^{\circ}$ comparing to the $n=2$ case, a 2D mirror MOT~\cite{wu_demonstration_2007} is likely preferred for optimal optical accesses. The ONF can be integrated on top of the mirror along $z$. The zero-field line can be shifted freely in 2D by adjusting, {\it e.g.}, the currents surrounding the plate pair on the left of Fig.~\ref{figField}d. 

\section{Experimental demonstration}\label{sec:exp0}
We take advantage of two-plate compactness (Fig.~\ref{figSetup}c)
to construct a planar ONF-2D-MOT interface, schematically illustrated in Fig.~\ref{figMOT}. We further demonstrate transient absorption spectroscopy measurement with a measurement repetition rate of $f_{\rm rep}=250~$kHz. 


\subsection{Experimental Setup}\label{sec:exp}

\subsubsection{Two-plate vacuum assembly}\label{sec:2-plateV}
The experimental setup is illustrated with Fig.~\ref{figMOT}a. A tapered nanofiber is fabricated with standard method ~\cite{Ward2014,Hoffman2014,Brambilla2004,Bashaiah2024}. The transmission of ONF is quite high, $T>85\%$. The fiber is epoxied to a titanium mount. The central $l_0\approx 5$~mm ONF section is centered to the mount, beneath which two pair of right-angle prisms are epoxied to the surface for the MOT-beam retro-reflections. The central pair of prisms are $10~$mm high.  The two side prisms are $10~$mm high too. The ONF mounting assembly, carefully chosen to be non-magnetic, is connected to a CF35 stainless steel (SS-304) conflate flange. The ONF is optically feedthroughed to the vacuum with a Teflon interfacing technique~\cite{Abraham1998}. The whole vacuum assembly is maintained to a $10^{-7}$~Pa base pressure. By opening a valve of a rubidium reservoir, a $^{87}$Rb partial pressure at  $10^{-6}$~Pa level is achieved near ONF. To avoid Rb condensation, we gently heat the titanium assembly to $30^{\circ}$C. Similarly, the ONF is heated by 100~$\rm {\bf \mu} W$ of $843~$nm light, (Fig.~\ref{figAbs}a) which is combined with the $780~$nm resonant probe beam with polarization optics and holographic gratings. 


The two-plate assembly is constructed outside the vacuum with plate dimensions of $L=120~\text{mm}$ and $W=40~\text{mm}$. As shown in Fig.~\ref{figPerm}, each ${\bf \mu}-$metal plate is magnetized by a pair of cylindrical NdFeB magnets (25~mm diameter, 10~mm height) positioned 60~mm apart along the $z$ axis. An inter-plate gap of $2h=70~\text{mm}$ is used, which exceeds the ideal condition $h \ll W$ due to the vacuum tube diameter. The relatively thick plates, with $d=10~\text{mm}$, are advantageous for the 2D-MOT operation distance of $h=35~\text{mm}$ in this ex-vacuo design. To allow dynamic adjustment of the magnetic field, each ${\bf \mu}-$metal plate is wrapped with 23 turns of insulated copper wire with 1~mm diameter. The current passing through one plate is subsequently referred to as the ``plate-coil'' current (see Fig.~\ref{figMOT}d). Additionally, a coil with diameter $D=280~\text{mm}$ is mounted 35~mm away to provide a $B_x$ bias field, with its current denoted as the ``x-coil'' current.
Simulation with Radia~\cite{Elleaume1997,Chubar1998} (Appendix~\ref{sec:simu}) confirms that the offset field generated by the $x-$coil is uniform across ONF to within $1\%$.

We characterize the field distribution with a combination of numerical simulation (Fig.~\ref{figField}(a,b)) and Gauss-probe measurements (Fig.~\ref{figPerm}). As detailed in Appendix~\ref{sec:simu}, by adjusting in the simulation the the NdFeB strength, the numerical results can match quite well with the measurements. The surface field near $S_2$ is highly uniform and stable (Fig.~\ref{figPerm}). For the two-plate assembly, we find a quasi-2D magnetic field-zero line near the ONF location with field gradients of $B_1^x\approx 21~{\rm G/cm}$, $B_1^y\approx -24~{\rm G/cm}$, $B_1^z\approx 3~{\rm G/cm}$ along $x,y,z$ respectively.


\subsubsection{Prism MOT}

As shown in Fig.~\ref{figMOT}a, a MOT beam of $18~\text{mm}$ diameter is directed through the vacuum viewport toward the central right-angle prism pair atop the titanium assembly. The laser beam, with power $P=20~\text{mW}$, is detuned by $\Delta = -2\Gamma$ ($-12~\text{MHz}$) from the $F=2 \rightarrow F'=3$ D2 hyperfine transition of $^{87}$Rb. It is circularly polarized to align with the quadrupole field direction, as depicted in Fig.~\ref{figSetup}c. Additionally, $1~\text{mW}$ of laser power resonant with the $F=1 \rightarrow F'=2$ hyperfine transition is combined with the MOT beam to provide hyperfine repumping. To prevent the ONF from encountering a dark line in the retro-reflected beam, caused by a small gap between the prisms, we slightly tilt the beam’s incidence angle.

To provide axial cooling and trapping, two additional $z$-beams are sent through the same viewport and redirected by the pair of smaller prisms to intersect at the ONF. These $z$-beams have a smaller diameter of $3~\text{mm}$, each with a power of $P=3~\text{mW}$. With a weak axial magnetic gradient $B_1^z \approx 3~\text{G/cm}$, we find the polarizations of the $z$-beams to be noncritical for 2D-MOT operation. Setting the $z$-beam circular polarizations in line with the field direction slightly compresses the MOT along the $z$-axis, as expected. However, in our setup, we intentionally choose the polarizations as opposite to the field direction, which helps to extend the 2D-MOT length by approximately $20\%$. A typical MOT image is shown in Fig.~\ref{figMOT}c, together with three mirror images.


\subsection{MOT loading}\label{sec:loading}
The MOT loading process is monitored by the ONF fluorescence. As illustrated with Fig.~\ref{figMOT}d setup, we direct an ONF output to a holographic grating to select the 780~nm fluorescence for single-photon counting (SPCM-AQRH, Excelitas). We switch off and back on the MOT beams with an acousto-optical modulator (AOM) to record the MOT-filling process. From the camera image, the final length of 2D-MOT along $z$ and in the $x-y$ plane are $l=4~$mm and $\Delta x=0.8~$mm respectively.

The fluorescence counting rate curve as shown in Fig.~\ref{figMOT}e includes a rapid initial growth which is followed by a slow increase further. The underlying mechanism is explained as following:  Magneto-optical trap is known to have an irregular density distribution~\cite{vengalattore_enhancement_2004} associated with the quite unavoidable variations of intensity-imbalance between counter-propagating MOT beams. During the initial MOT loading, atoms prefer to be accumulated in intensity-balanced locations. Later, when there are plenty of atoms in these locations, multi-scattering starts to drive their expansions, leading to a large, ``density-limited '' sample~\cite{Townsend1995} with a uniform local density $\rho_0$. Here, the rapid initially growth of ONF fluorescence is associated with filling of intensity-balanced spots across ONF. The slow increase of ONF fluorescence later is associated with the multi-scattering-driven spreading to uniformly fill the ONF interaction zone, which effectively increase the interaction length $l$. We fit the fluorescence curve with a phenomenological model to include this multi-scattering-driven MOT expansion dynamics, detailed in Appendix~\ref{secFit}. The multi-scattering-limited density $\rho_0$ is derived from  ONF absorption measurement. In particular, we find that the fully loaded atomic sample  attenuates the ONF-guided resonant probe with an ${\rm OD}=2.2$ optical depth (see {\it e.g.} Fig.~\ref{figAbs}b), which suggests a uniform $\rho_0 = 3\times10^{10}$/cm$^3$ over the $l=4~$mm interaction length.
In Fig.~\ref{figMOT}e we also estimate $N_{\rm eff}={\rm OD}(t)/{\rm od}_1$ as an effective number of atoms coupled to ONF. Here ${\rm OD}(t)$ is scaled with the fluorescence counting rate during the MOT loading. The ${\rm od}_1=0.09$ is single-atom resonant optical depth at a $125~$nm distance away from the $d=500~$nm ONF~\cite{Solano2017,Ma2023} in our setup (see Appendix~\ref{secFit}). 

\subsection{Field control}

We shift the 2D-MOT along \( x \) and \( y \) by adjusting the \( x \)-coil and plate-coil currents, demonstrating this field control by measuring the ONF-coupled steady-state MOT fluorescence, as shown in Fig.~\ref{figMOT}f.

We begin by optimizing the ONF fluorescence counting rate through adjustments of both the \( x \)-coil and plate-coil currents. With the plate-coil current then fixed at its optimal 3.5 A, we scan the \( x \)-coil current in small steps between 0.76 A and 1.5 A. This procedure shifts the 2D-MOT along \( x \) by approximately 1 mm, as indicated by the mirror images observed on the camera (upper-most and lower-most MOT images in Fig.~\ref{figMOT}c). Since the 2D-MOT has a transverse size of about 0.8 mm, this shift is sufficient to toggle the ONF-coupled fluorescence off and back on, as shown by the green markers in Fig.~\ref{figMOT}f.

Next, with the \( x \)-coil current set to its optimal 1.16~A, we scan the plate-coil current between 1.2~A and 5.4~A. This adjustment shifts both the 2D-MOT and its second-order image transversely by approximately 1.45~mm on the camera (the two central images in Fig.~\ref{figMOT}c). The resulting variations in the ONF fluorescence counting rate are represented by the red markers in Fig.~\ref{figMOT}f.

From the data in Fig.~\ref{figMOT}f, we estimate the plate-coil current tuning response to be approximately 0.8~G/A at the 2D-MOT location. Notably, the field is enhanced by a factor of \( \eta \approx 4 \) due to the magnetization of the \( {\bf \mu} \)-metal. Without the \( {\bf \mu} \)-metal, the same plate-coil would produce only about 0.2 Gauss per Amp at the same distance. This soft-magnetic field enhancement leads to the low power operation. In this setup, \( \eta \approx 4 \) is constrained by the moderate \( d/W \) ratio and the resultant strong depolarization field~\cite{JacksonBook}. In future work, a shorter working distance \( h \) and a thinner plate with a smaller \( d/W \) ratio could reduce demagnetization, potentially increasing  \( \eta \)  to the hundred level. Regardless of the $\eta$ value, the soft-magnetization protects field uniformity against the plate-coil winding imperfection, as being explained in Sec.~\ref{sec:uniform}. 


\subsection{High-rep-rate transient ONF spectroscopy}\label{sec:specH}
\begin{figure*}[htbp]
        \centering
        \includegraphics [width=1\linewidth]{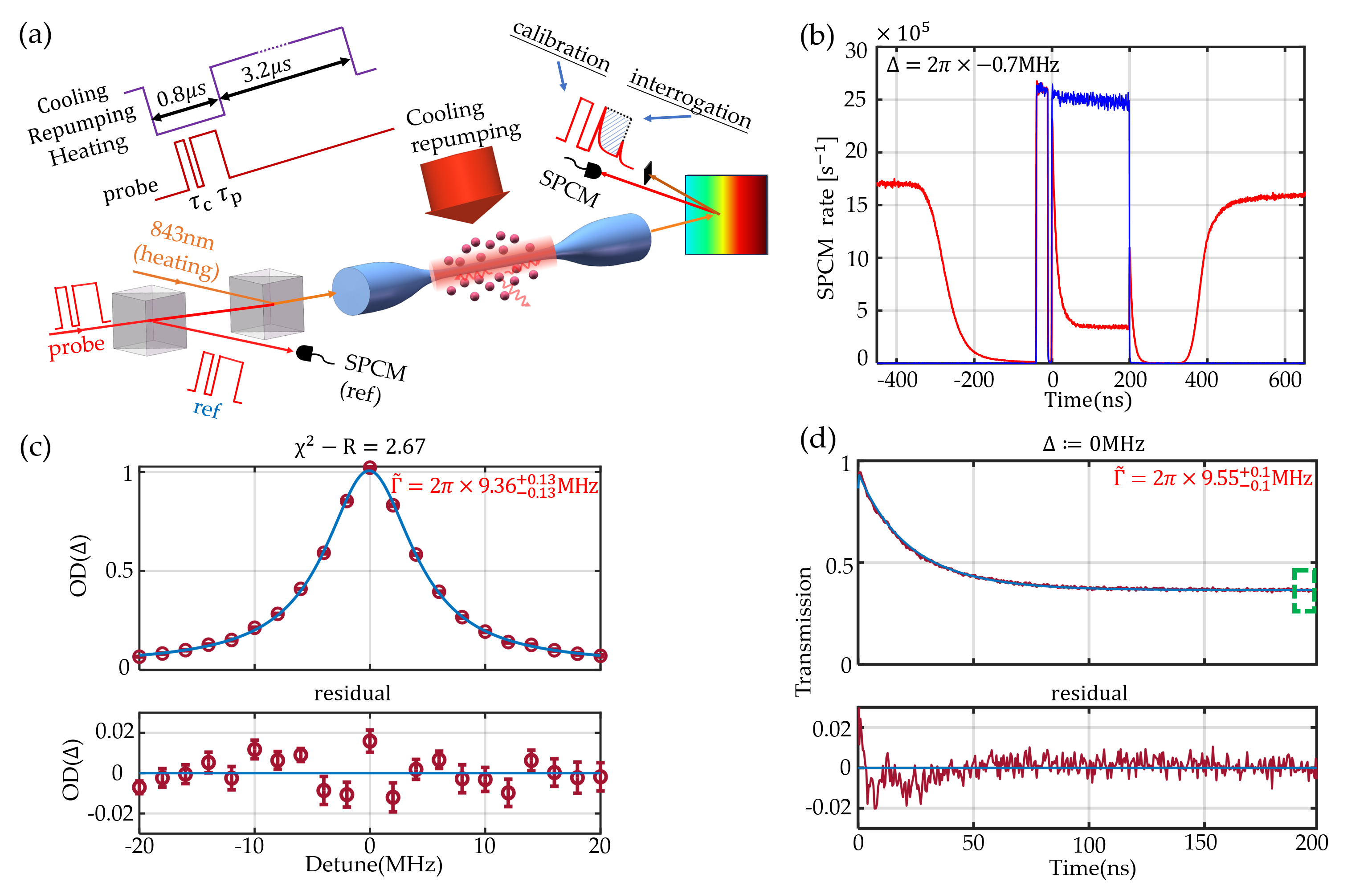}
     \caption{(a): Absorption spectroscopy setup and timing sequence. SPCM: Single photon counting module. Here $\tau_{\rm c}=30~$ns. $\tau_{\rm p}=200~$ns in (b) and (c,d) respectively. (b): Typical SPCM readout of the ONF-transmitted probe (red) and reference (blue). (c): Steady-state absorption spectroscopy and Lorentzian fit.  (d): Transmission of a resonant probe ($\Delta=0$) from (b) with fit according to linear dispersion theory ~\cite{Chalony2011,Kwong2014,JacksonBook,Su2023b} (Appendix~\ref{sec:lwApp}). The dashed box suggests the range of data for the average to compile the $\Delta=0$ data point in (c).}
      \label{figAbs} 
 \end{figure*}

As illustrated in Fig.~\ref{figField}(a,b), even with a working distance of \( h=35 \)~mm, the field distribution in the two-plate configuration remains highly elongated along \( z \). The 3~Gauss/cm gradient leads to 0.6~Gauss field strength deviation across the \( l=4 \)~mm ONF interaction section aligned to the quasi-zero-field line. This level of field suppression should be sufficient for atomic spectroscopy at moderate precision, allowing measurements without globally switching off the field.

Leveraging this near-zero-field feature, we design a cooling-probe sequence for high-speed atomic spectroscopy, as depicted in Fig.~\ref{figAbs}a, simply by modulating the laser beams. The measurement repetition period is \( T_{\rm rep} = 4~{\bf \mu} \)s. Within each period, the cooling and repumping beams, along with the 843~nm ONF heating beam, are all turned off and then back on within 0.8~\( {\bf \mu} \)s. As in Fig.~\ref{figAbs}b, accounting for finite AOM switching time, \( T_{\rm p} \approx 600 \)~ns is open for absorption measurements.
This probe window is short enough to prevent significant atomic motion, enabling atoms to be recaptured by a 3.2~\( {\bf \mu} \)s cooling pulse for the next probe cycle. 

During the probe window, we generate a composite probe pulse using a fiber electro-optical modulator (fEOM)-based optical arbitrary waveform generator (OAWG)~\cite{He2020c}. The pulse includes a 30~ns ``calibration pulse'' followed by a $\sim 200$~ns ``interrogation pulse''. The calibration pulse is 400 MHz blue-detuned from the \( F=2 \rightarrow F'=3 \) hyperfine transition, leaving its ONF transmission unaffected by the presence of cold atoms. The interrogation pulse is scanned in detuning \( \Delta \) from -20 MHz to 20 MHz. We keep the probe power at a low level of 5~pW to avoid saturating the atomic transition~\cite{foot:sat}. For each detuning \( \Delta \), we integrate the single-photon counting module (SPCM) measurements over \( T_i=60 \) seconds. The resulting single-photon counts from all \( f_{\rm rep} \times T_i = 1.5 \times 10^7 \) measurements are histogrammed to yield the ONF-transmitted probe signal \( I_p(t) \), shown in red in Fig.~\ref{figAbs}b. We note the MOT fluorescence is visible in \( I_p(t) \) beyond the probe time window, as expected.

Additionally, a portion of the probe pulse is diverted to a reference SPCM without passing through the ONF, providing a ``reference'' signal \( I_r(t) \). By comparing the 30 ns ``calibration pulse'' between the probe and reference signals, we normalize \( I_r(t) \) (displayed in blue in Fig.~\ref{figAbs}b). With the normalization, the time-dependent probe transmission during the interrogation time is obtained as \( T(t) = I_p(t)/I_r(t) \) in Fig.~\ref{figAbs}d.

\subsection{Linewidth measurement}\label{sec:spec}
With the transient transmission data $T(t)$ as those in Fig.~\ref{figAbs}d, we retrieve the near-field coupled atomic response in two ways (Appendix~\ref{sec:lwApp}). First, as suggested by the green dashed box, we take the $T(t)$ data between $195~{\rm ns}<t<200~{\rm ns}$ and calculate the average $\bar T$ at the steady state. We then evaluate the optical depth of the evanescently coupled cold atoms according to Beer-Lambert law~\cite{Asenjo-Garcia2017}, ${\rm OD}=-{\rm ln}(\bar T)$, at each probe detuning $\Delta$. Typical absorption spectroscopy data are given in Fig.~\ref{figAbs}c together with Lorentzian fit. In the second method, we choose a particular $T(\Delta,t)$ curve as those in Fig.~\ref{figAbs}b, and apply linear dispersion theory~\cite{Chalony2011,Kwong2014,JacksonBook,Su2023b,CardenasLopez2023} to fit the Lorentzian atomic response.  In both cases, we arrive at a fitted linewidth of $\tilde \Gamma/ 2\pi \approx 9.5\sim 10.5~$MHz.  This value is substantially larger than one might expect from the $\Gamma=2\pi\times 6.1~$MHz natural linewidth of the D2 transition~\cite{Steck2003}, even after accounting for the ONF-enhanced emission and the surface interactions~\cite{Goban2012a,Patterson2018,Solano2019c}. 

It is worth noting that unlike previous spectroscopic work with free atoms~\cite{Sague2007,Patterson2018, Solano2019c}, where a prolonged measurement could affect the atomic trajectories near the nanofiber, here the atomic motion is effectively frozen for our nanosecond probe. As we expect a largely uniform atomic density in the near field, the spectroscopic measurement can be quite conveniently interpreted to infer surface interactions~\cite{Peyrot2019a}. Therefore, that we see a Lorenzian absorption profile as in Fig.~\ref{figAbs}(c,d), broadened but without substantial van-der-Waals-shift induced asymmetry~\cite{Sague2007,Nayak2012, Patterson2018, Solano2019c}, is quite puzzling. In Appendix~\ref{sec:lwApp} we detail the Fig.~\ref{figAbs}(c,d) measurements, and provide strong evidence that the anomalously large linewidth as in Fig.~\ref{figAbs}(c,d) is affected by the surface quality of ONF~\cite{Nayak2012} and is likely of magnetic origin.

Here, we note that the observed linewidth broadening can hardly be attributed to the residual field of our two-plate assembly alone. In particular, when the ONF is aligned along the field-zero line, the measured gradient of $B_1^z \approx 3$~G/cm should lead to a field variation of only $\delta B = 0.6$~G. The resulting average field strength would therefore remain at the 0.3~G level, far below the required field to broaden the line by even 1~MHz (approximately 1~G), according to numerical simulations~\cite{Qiu2022,Ma2023} (Appendix~\ref{sec:mag}). Instead, explaining the observed broadening would require $B_1^z \approx 15$~G/cm, which is clearly inconsistent with the 2D-MOT observations in Sec.~\ref{sec:MOT}.

\section{Toward an ultra-straight 2D ferromagnetic trap}\label{sec:straight}

As shown in Fig.~\ref{figField}(a,b), the field zero in the two-plate assembly is inherently three-dimensional. The field variation along $z$ only vanishes in the limit $\varepsilon = h/L \ll 1$. On the other hand, this field variation is absent in the four-plate configuration, as described in Fig.~\ref{figField}(c,d). Of course, this ideal behavior assumes that the ${\bf \mu}$-metal plates are perfectly magnetized and precisely aligned. In this section, we examine practical aspects of this assumption by measuring the field distribution and comparing it with numerical simulations.

Before proceeding with the discussion, it is worth noting that the highly uniform field distribution shown in Fig.~\ref{figField}(c,d) arises from the symmetry of the source field, which is robustly enforced by the magnetic response (Sec.~\ref{sec:uniform}) of the precisely aligned $\mu-$metal plates. In principle, comparable uniformity could be achieved using current-carrying wires alone~\cite{foot:coils}. In practice, however, achieving this is technically challenging: aside from the difficulty of perfect winding, the current flow within conductors does not always conform precisely to the wire geometry~\cite{Wang2004,Aigner2008}. 

\subsection{Field uniformity and stability}\label{sec:perm}

\begin{figure}[htbp]
    \centering
    \includegraphics [width=1\linewidth]{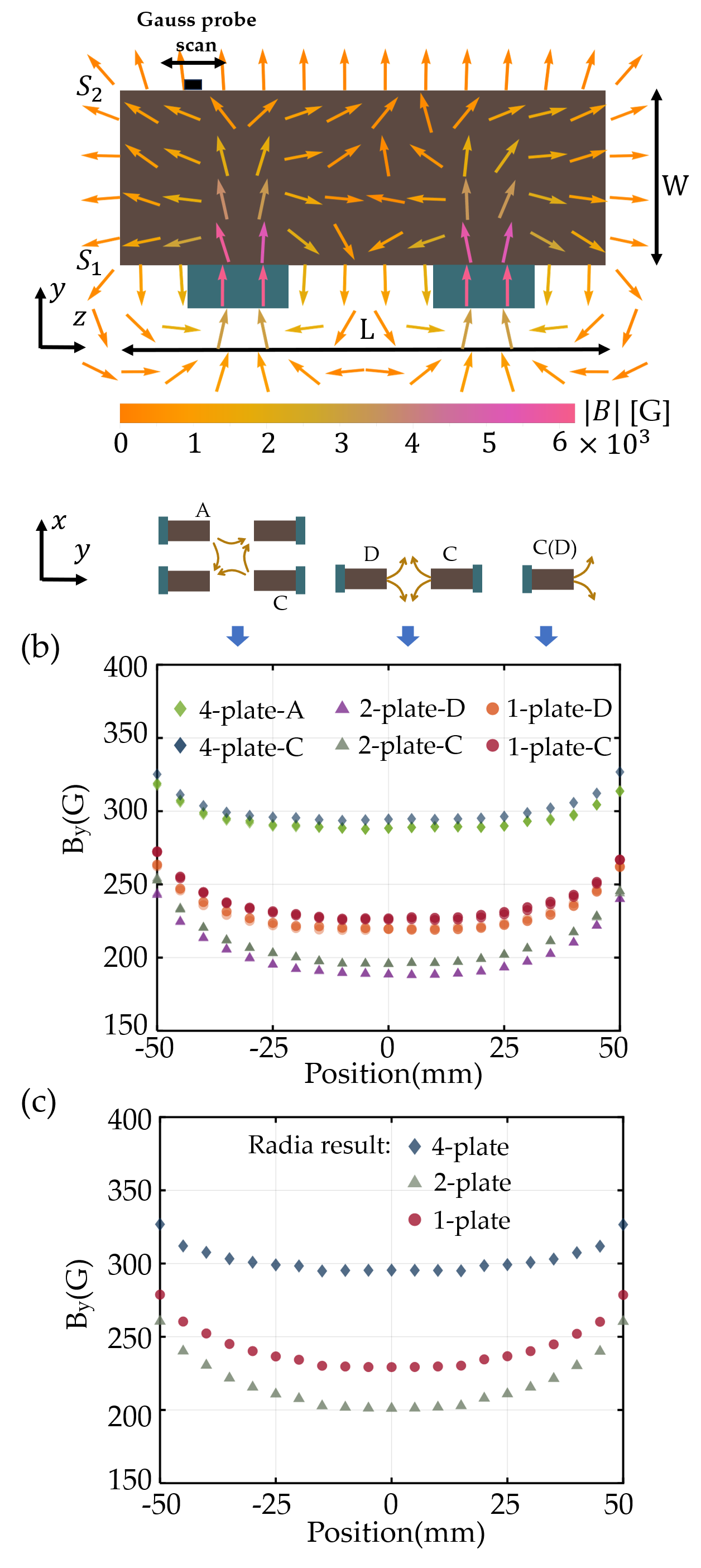}
     \caption{(a): Field distribution of the NdFeB$-{\bf \mu}-$metal structure in this work, obtained numerically with Radia similar to Fig.~\ref{figSetup}b. The Gauss probe for the scanning measurement is suggested on top of the $S_2$ surface. (b): The surface magnetic field for two of the plates in the 4-plate assembly (diamond symbols),  2-plate assembly (triangular symbols), and two individual plates (disk symbols). The plate-assembly configurations are illustrated on the top. (c): Simulated surface fields using Radia~\cite{Elleaume1997,Chubar1998} (Appendix~\ref{sec:simu}).}
      \label{figPerm} 
 \end{figure}
 
With in mind the field uniformity in the $n$-plate assembly is supported by the field uniformity generated by each plate, in this section, we focus on the $S_2$ surface field measurements for individual plates in the 4-plate (Fig.~\ref{figField}(c,d)), 2-plate (Fig.~\ref{figField}(a,b)), and single-plate (Fig.~\ref{figPerm}(a)) setups. 
For the purpose, we construct a prototype 4-plate assembly according to Fig.~\ref{figField}(c,d), with $h=35~$mm and $h'=17.5~$mm. All the NdFeB$-{\bf \mu}-$metal structures in the assembly share the same size and shape as those in Sec.~\ref{sec:2-plateV}. The ${\bf \mu}-$metal plates are milled in shape with $10~{\bf \mu}$m precision. The positions of NdFeB blocks and the ${\bf \mu}-$metal plates are adjusted according to a millimeter-scaled ruler, and is therefore with quite moderate precision at 0.1~mm level. Then, as illustrated in Fig.~\ref{figPerm}(b) on the top, by detaching the NdFeB blocks from plate-A and plate-B and reversing the NdFeB blocks on plate-D, we effectively recreate the 2-plate assembly according to Fig.~\ref{figField}(a,b). Finally, by detaching the NdFeB blocks for either plate-C or plate-D, field distribution for individually magnetized single NdFeB$-{\bf \mu}-$metal structure is created near the respective $S_2$ surface.   There are appreciable differences in the surface field strength in these configurations, as in the Fig.~\ref{figPerm}(b) measurements to be discussed shortly, due to the mutual magnetization.

The field measurement is performed with a Gauss probe with 1~mm sensor size and $0.5~$G resolution. The Gauss probe is mounted on a 3D translation stage, centered to the $S_2$ surface of the plate to be measured, and is gently pressed against the surface to fix the distance. Then, as in Fig.~\ref{figPerm}a, by translating the Gauss probe along $z$, the surface field is recorded across the $L=120~$mm plate length. The measurements are repeated over days to verify the field stability. Example data are given in Fig.~\ref{figPerm}b with diamond, triangular, and disk markers. The marker size is chosen according to the measurement uncertainty and our confidence in positioning the Gauss probe.

We first note that all the $B_y(z)$ field profiles in Fig.~\ref{figPerm}b align well with the expectations from numerical simulations~\cite{Elleaume1997,Chubar1998} shown in Fig.~\ref{figPerm}c (Appendix~\ref{sec:simu}). Slight but repeatable deviations in $B_y$ are observed on a $\sim 10~\mathrm{mm}$ scale, with a magnitude close to the 0.5~G resolution limit. These deviations are likely caused by corrugations along the edges of the $d = 10~\mathrm{mm}$ $S_2$ surface, introduced during the chamfering process in these prototypes. Such effects can be mitigated in the future with improved machining techniques.

The measurement results are repeatable. Taking the 1-plate-C data as an example, which consists of three repeated measurements conducted over the span of a week, the data deviation is less than 1~G, showing excellent overlap in the plot. On the other hand, given that NdFeB blocks exhibit variations in surface quality and magnetic strength, the overall surface field strength naturally varies from plate to plate, as illustrated in the examples in Fig.~\ref{figPerm}b. These variations can be managed by hand-picking NdFeB blocks with consistent properties or by fine-tuning the field strength using additional magnets.

We verify that the uniform field profile shown in Fig.~\ref{figPerm}b is insensitive to the way the source magnets are attached on the $S_1$ surface. Field variations at the $1\%$ level only occur near $S_2$ when one of the two NdFeB blocks is displaced on the $S_1$ surface by as much as a few millimeters—a displacement much larger than the positioning precision achievable with a millimeter-scale ruler. As discussed in Sec.~\ref{sec:uniform}, the combination of efficient source-field shielding and surface field energy minimization enforces a uniform field distribution near $S_2$. For plates with large enough $W$, $L$ and $d$, field irregularities induced by variations of small magnets on the distant surface are suppressed. Similar shielding effects are expected to smooth out the irregularity in the plate-coil current~\cite{wu_demonstration_2007}.

Given the highly uniform surface field in the finely machined 4-plate assembly, the field distribution above $S_2$ for $h,h' \ll L$  are also expected to be uniform (Fig.~\ref{figField}(c,d)), as dictated by field continuity~\cite{foot:closure}. In our prototype assembly, accurately mapping the 3D field distribution is technically challenging. Nevertheless, using the same Gauss probe, we measured the field both above the plate surfaces and near the zero-field region of the 4-plate assembly. These measurements align with simulations, albeit with reduced fractional resolution and increased positional uncertainties.

As in Sec.~\ref{sec:uniform}, we emphasize the choice of large enough working distances $h$ for finely machined ${\bf \mu}-$metal plates, to avoid fringe fields associated with micrometer-scale ${\bf \mu}-$metal surface corrugations~\cite{Wang2004, ThesisWu}.

\subsection{Alignment considerations}
The construction of the 2D quadrupole field shown in Fig.~\ref{figField}(c,d) relies on precise alignment of the four uniformly magnetized \( {\bf \mu} \)-metal plates. In practice, field deviations can arise from misalignment or positional errors. However, because the \( {\bf \mu} \)-metal plates can be accurately machined and securely mounted, we anticipate small alignment or positional errors in the plate assembly will be manageable to induce negligible field deviations.


First, when the four plates are not perfectly parallel, then, depending on the kind of relative angular misalignment, the zero-field line is affected in different ways. However, with the rigidity provided by the precisely machined ${\bf \mu}-$metal plates, we estimate that angular alignment precision can reach a 10-micro-radian level. For 2D trap with moderate field gradients, {\it e.g.} $B_1 \sim  100~\text{G/cm}$, the residual field at center within a $l=~$100~mm distance due to the angular errors would be limited to 10~milli-Gauss level, regardless of relative orientations among the plates. 

Next, we consider the effect of a positional error $\delta z$ along the $z$-axis by one of the plates. The resulting field perturbation can be interpreted as a shift in surface magnetic charge~\cite{JacksonBook}. The unbalanced magnetic charge at the edges leads to a residual field at the center, which, for sufficiently large plate length $L$,  becomes weak and smooth. Therefore, for precisely positioned plates with $\delta z/L$ at {\it e.g.} $10^{-4}$ level, and in addition $L\gg h$, this type of perturbation can be suppressed to milli-Gauss level while maintaining the $B_1\sim 100~$G/cm gradient.

On the other hand, for any one of the plates, small positional errors along the $x$ and $y$ directions shifts the zero-field line itself, and can be compensated by fine-tuning the magnetization through adjustments to the plate-coil current.
\subsection{Field-free ONF interface with $l\sim 100~$mm}

So far, we have shown that it is practical to construct a 2D ferromagnetic trap as those in Fig.~\ref{figField}(c,d) with high precision. Enhanced by permanent magnets, a field gradient of $B_1=100~$G/cm can be quite easily achieved at $h=20~$mm distance for a four-plate assembly with {\it e.g.} $L=200$~mm. The center zero-field line is long enough to support ONF-2D-MOT system with $l\sim 100~$mm, with residual field suppressed to the milli-Gauss level. As such, the evanescent optical lattice~\cite{Vetsch2010,Goban2012a,Meng2018,Su2023b,Kestler2023} can be loaded from the high-gradient 2D-MOT without switching off the trapping field.   

For a uniformly filled lattice, the collective interaction strength between guided photons with the lattice atoms increases linearly with $l$. Scaling up from an existing apparatus~\cite{Su2023b}, with the $l\approx 100~$mm interaction length,  the resonant optical depth for an atomic array would reaches ${\rm OD}\sim 10^3$ level. This interaction strength would be quite unprecedented, which can be crucial for uncovering new regime in waveguide QED~\cite{Asenjo-Garcia2017,Solano2017}. Importantly, the quantum optical measurements can be performed in a quasi-continuous manner, as those demonstrated in this work (Fig.~\ref{figAbs}), leading to high data rate. The efficient laser-cooling in between measurements should also help to mitigate ONF vibration induced heating~\cite{Hummer2019}.



\section{Discussions}\label{sec:sum}

\subsection{Magnetic guiding}\label{sec:guide}
So far, with nanosecond ONF transmission spectroscopy for the $n=2$ assembly (Sec.~\ref{sec:specH}) and a field-characterization study for the $n=4$ version (Sec.~\ref{sec:straight}), our discussions have been focused on how the ultra-smooth field generated by the ferromagnetic assembly improves the ONF interface technology for  free~\cite{Patterson2018, Solano2019c} and lattice-confined atoms~\cite{Vetsch2010,Goban2012a,Meng2018,Su2023b,Kestler2023}. 

Clearly, with reduced $h$, the field gradient in these ferromagnetic traps can easily reach $B_1=100$~G/cm or higher, strong enough for tightly confining laser-cooled atoms in their low-field-seeking states with magnetic dipole forces~\cite{wu_demonstration_2007}.  Spin flips near the zero-field line may be suppressed by the vector shift of ONF guided light in the near field. Alternatively, taking advantage of the fast $\mu-$metal response~\cite{coey2009Book}, one may create a two-dimensional time-orbiting potential trap~\cite{Petrich1995}, by periodically driving the offset currents at high enough frequencies {\it e.g.} near 100~kHz. The setup may open up new quantum optical scenarios featuring waveguided photons interacting with magnetically guided atoms. To this end, we note related progresses are attempted with optical guiding inside hollow-core fibers~\cite{Xin2018a,Song2024}, where the field inhomogenuity poses the key challenge. Here, the ultra-smooth confining potential~\cite{wu_demonstration_2007} offered by the ferromagnetic trap would help to support matterwave coherence over a long distance comparable to that for the optical coherence in the ONF-based quantum optical platform~\cite{Solano2017, Solano2017b}.

\subsection{Beyond $n=4$: Multipole fields}
\begin{figure}[htbp]
    \centering
    \includegraphics [width=1\linewidth]{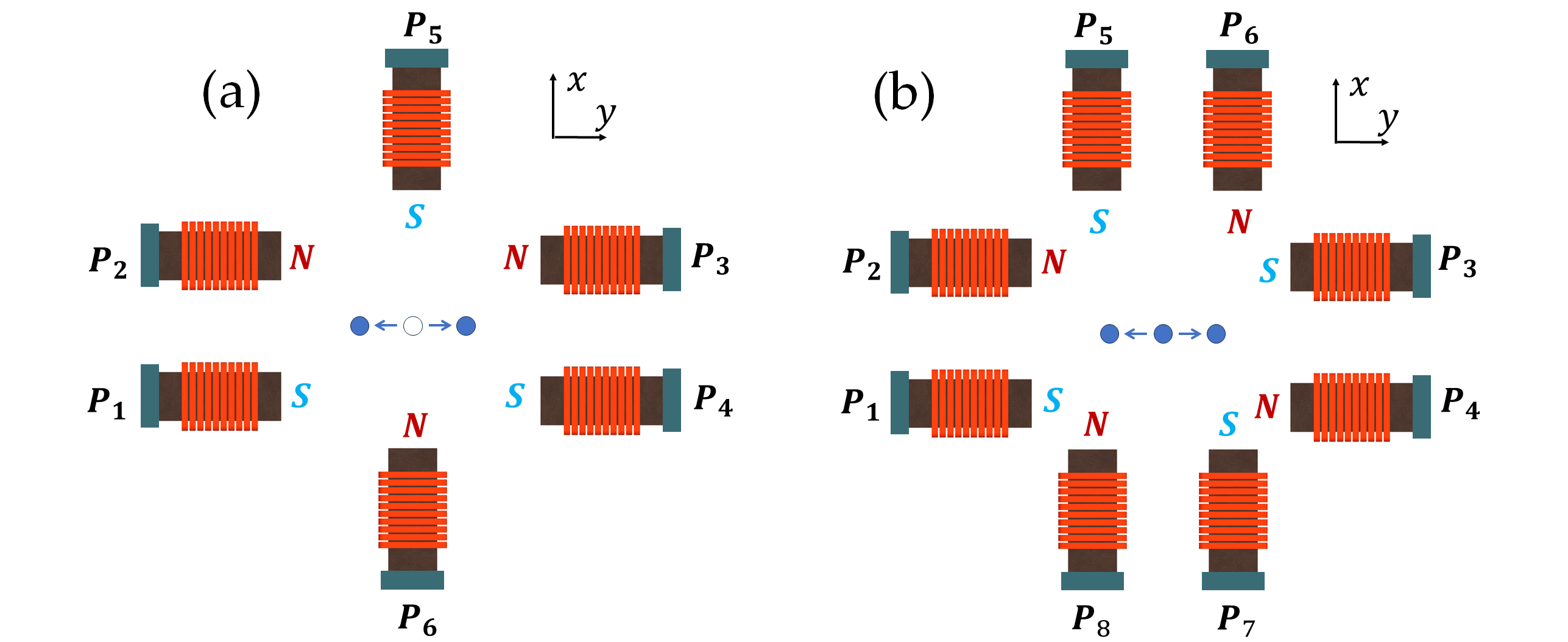}
     \caption{(a) Schematic of an $n=6$ assembly. The central open circle represents the hexapole field when the magnetization of the $P_{5,6}$ plates are set at a critical value, beyond which the second-order field zero is split into two two first-order zeroes, moving toward $P_{1,2}$ and $P_{3,4}$, as represented by the blue disks.  (b)  Schematic of an $n=8$ assembly which creates an octupole field when $P_1$ to $P_8$ share the same strength. By ramping up $P_{5\sim8}$ currents to increase their magnetization, the 3rd-order zero at center is split into three 1st-order zeroes. The splitting and merging of the multiple zero-field lines enables coherent control of multiple 1D atomic samples around ONF. }
      \label{figHO} 
 \end{figure}

The soft ferromagnetic assembly can be conveniently constructed to support various atomic cooling and trapping geometries~\cite{Wu2004, wu_demonstration_2007}. In particular, the linear quadruple field  in Fig.~\ref{figField} can be extended to higher-order trapping fields, such as the $n=6$ hexapole (Fig.~\ref{figHO}a) and $n=8$ octupole (Fig.~\ref{figHO}b) configurations. Similar to the $n=4$ configuration, the higher-order fields are generated by $\nu=n/2 \geq 3$ pairs of polarization-balanced magnets. The field variation along $z$ can therefore be efficiently suppressed to support highly uniform zero-field lines at the assembly center, as in Sec.~\ref{sec:straight}. These high-order fields can be quite useful to extend the functionality of the ONF interface. For example, since the field strength around the higher-order field zero scales as $|B(\rho)|\propto \rho^
{\nu-1}$ for $\rho=\sqrt{x^2+y^2}$, a higher-order trap provides a larger field-free volume~\cite{Andresen2010} around ONF during laser cooling. 

Furthermore, as schematically illustrated in Fig.~\ref{figHO}, by adjusting the relative strengths of ferromagnetic plate pairs, the high-order zero-field line can be split into multiple low-order zero-field lines. Taking the hexapole field in Fig.~\ref{figHO}a as an example. Here, the $P_{5,6}$-plates must be poled with sufficient strength to counter the $B_x$ field  at the assembly center generated by the $P_{1,2,3,4}$ plates. A hexapole zero is formed only at a critical $P_{5,6}$ poling strength, beyond which the 2nd-order zero is split into two 1st-order zeroes to the left and right. Similarly, for the octupole field in Fig.~\ref{figHO}b, by ramping up the poling strength of $P_{5,6,7,8}$ plates, the 3rd-order zero at the center is split into three first-order zeroes. The ability to split and merge multiple zero-field lines, combined with the ultra-smooth magnetic guiding (Sec.~\ref{sec:guide}), may enable coherent control of multiple 1D samples around ONF to interact with the guided photons.


\subsection{Summary and outlook}

Integrating nanophotonics with cold atoms represents a promising route toward quantum information processing via nonlinear quantum optics~\cite{Chang2014,Sheremet2023}. A major technical challenge in this endeavor is the precise positioning of cold atoms at the nanophotonic interface. Among the most successful approaches to address this challenge is the ONF-cold atom interface, which has seen significant progress over the last decade~\cite{Vetsch2010,Corzo2019,Solano2019c,Solano2017,Solano2017b,Patterson2018,Liedl2024, Pache2025}. In these works, sub-Doppler cooling is critical for efficient loading of the near-field lattice, while a uniform atomic spectroscopic response is essential to support long-range resonant dipole interactions. But the associated field-free requirement is at odds with the necessity of a strong field gradient for magneto-optical trapping. Managing these conflicting requirements becomes increasingly difficult at large $l$, the ONF interaction distance, which is preferred to be as long as possible. But switching off the field to accommodate these constraints severely limits the duty cycle of single-photon-level measurements.


In this work, we have proposed to resolve the contradictory field requirement at the cold-atom-ONF platform using 2D ferromagnetic trap~\cite{vengalattore_enhancement_2004,wu_demonstration_2007}. We highlight the unique opportunity of aligning the zero-field line in 2D-MOT with the stiff ONF, for efficient loading and field-free operation of the nanophotonic interface. We show that precisely fabricated ${\bf \mu}-$metal plates robustly uniformize source field from permanent magnets or current-carrying wires, for generating ultra-straight 2D field to match a long and straight ONF~\cite{Zhang2024}. 

Experimentally, we integrate a two-plate ferromagnetic assembly with ONF using a planar 2D-MOT (Fig.~\ref{figMOT}). The field uniformity is confirmed with Gauss probe measurement on the plate surface (Fig.~\ref{figPerm}b). The 2D nature of the central quadrupole field (Fig.~\ref{figField}(a,b)) is verified by the robust operation of the highly elongated 2D MOT (Fig.~\ref{figMOT}e). Freed from the slow field-switching constraints, we perform atomic spectroscopy at a repetition rate as high as $f_{\rm rep} = 250~$kHz, utilizing nanosecond probe pulses as shown in Fig.~\ref{figAbs}.  We note that the quasi-continuous spectroscopic measurements minimally perturb the atomic trajectories~\cite{Patterson2018, Solano2019c}. In the ONF near field, the atomic density distribution is preserved as that during MOT operation, likely to be quite uniform. Therefore, our ONF spectroscopy should reflect the near-field shifts in a straightforward manner. As in Sec.~\ref{sec:spec} and detailed in Appendix~\ref{sec:lwApp}, that the absorption profile is uniformly broadened without notable asymmetry~\cite{Nayak2012} is quite unexpected and warrants further investigation. 

The $\mu-$metal soft ferromagnetics are excellent field uniformizers, and can be precisely machined and positioned to generate exceptionally accurate 2D fields. In Sec.~\ref{sec:straight} 
we provide arguments that a field-free ONF interface with $B_1\sim 100$~G/cm and $l\approx 100~$mm can be constructed using an $n=4$ assembly, where an ${\rm OD}\sim 10^3$ may be reached for an ONF-lattice~\cite{Su2023b} with 2D-MOT loading. The quantum optical measurements can be performed in a quasi-continuous manner, as those demonstrated in this work. The efficient laser-cooling in between measurements may help to mitigate ONF vibration induced heating~\cite{Hummer2019}. The unprecedentedly high light-atom coupling strength may help to unlock new research opportunities in waveguide QED~\cite{Asenjo-Garcia2017,Solano2017, Sheremet2023}. Bringing the ferromagnetic plates closer to the center, even higher gradients at $B_1\sim$~1~kG/cm level appears achievable, where the ONF may serve as a local probe to study cooling-trapping dynamics in a high-gradient 2D-MOT~\cite{Haubrich1996}. Additional cooling in the high-gradient magnetic trap should lead to tightly confined 1D atomic samples, with which a natural step forward is
to combine the ONF interface with magnetically guided atom interferometry~\cite{wu_demonstration_2007} to explore controllable quantum interaction between the waveguided photons and atoms.





\section*{Acknowledgments}
We are grateful to Professor Luis Orozco and Professor Wei Fang for valuable discussions. This project is supported by  National Key Research Program of China under Grant No. 2022YFA1404204, from Natural Science Foundation
 of China under Grant No.~12074083,~12274272, from Natural Science Foundation of  Shanghai
under Grant No. 23dz2260100, and from the Shanghai Science and Technology Innovation Action Plan  under the Grant No. 24LZ1400300.

\section*{DATA AVAILABILITY}
 Data and simulation codes underlying the results presented in this paper are not publicly available at this time
 but may be obtained from the authors upon reasonable
 request.

\appendix

\section{Magnetic field simulation}\label{sec:simu}
We use the Radia program~\cite{Elleaume1997,Chubar1998}, a free magnetic field simulation software based on \textit{Mathematica}, to simulate the magnetic field of the $n$-plate assembly. The field distribution in Radia is determined by evaluating the magnetization of each NdFeB$-{\rm {\bf \mu}}$-metal plate. For this purpose, the (soft) magnets are segmented into sufficiently small units to construct an interaction matrix~\cite{Elleaume1997,Chubar1998}. A relaxation algorithm is then applied to determine the stable magnetization.

In the simulation, the ${\rm {\bf \mu}}$-metal parameters are chosen as ${\bf \mu}_r=10^5$, $B_{\rm sat}=0.8~\mathrm{T}$, and $B_{\rm rem}=0$~\cite{coey2009Book}. The surface field strength of the NdFeB block is set to $B_0=0.9~\mathrm{T}$, an empirical value selected to align the central field gradient of the assembly between measurements and simulations. The structural geometry corresponds to the experimental setups shown in Fig.~\ref{figField} and Fig.~\ref{figPerm}. 

The accuracy of the field simulation relies on the granularity of magnet segmentation and the numerical relaxation tolerance. We adjusted these parameters to ensure that neither finer subdivisions nor tighter relaxation tolerances significantly affected the final results. The relaxation tolerance is set to ${\rm prec}=10^{-4}$~T (1~G). For the simulations in Fig.~\ref{figField} and Fig.~\ref{figPerm}c, each NdFeB block is divided into $10\times10\times10$ subunits. Each ${\rm {\bf \mu}}$-metal plate is divided into $n_z=160$ and $n_y=50$ segments along the $z$ (``$L$ direction'') and $y$ (``$W$ direction'') axes, respectively. Additional simulations confirmed that the central field profile $B_y(z)$, as shown in Fig.~\ref{figPerm}c, is insensitive to segmentation along the $x$-axis, allowing us to set $n_x=1$ to conserve computational resources. This approach neglects the variation of $B_y$ across the plate thickness ($x$), which is minimized at the center ($x=0$) and peaks at the edges. To account for this approximation, an overall scaling factor of 0.8 is applied to the field simulation in Fig.~\ref{figPerm}c, evaluated at $h=0.5~\mathrm{mm}$ from $S_2$, to match experimental measurements shown in Fig.~\ref{figPerm}b. This correction is necessary only in the vicinity of $S_2$. For field distributions where $h \gg d$, the simulation remains accurate, and we fix the NdFeB block surface field at $B_0=0.9~\mathrm{T}$ for all simulations.

\section{A multi-scattering model for loading the 2D-MOT-ONF interface}\label{secFit}

The process of loading a magneto-optical trap is characterized by the time-dependent atom number and the characteristic size of the MOT, $N(t)$ and $\sigma(t)$ respectively. For a single-chamber MOT where atoms are directly captured from the background vapor, as in this work, the $N(t)$ dynamics follows quite accurately the differential equation $\dot N=R-\kappa N-\beta N^2/\sigma^3$. Here the loading rate $R$ depends on the laser and magnetic field parameters, as well as the background rubidium partial pressure $P_{\rm Rb}$. Instead, the single-body loss rate $\kappa$ is almost solely decided by the total pressure $P$. The two-body loss coefficient $\beta$ describes the light-assisted collisional losses~\cite{prentiss1988}. For the setup in this work (Fig.~\ref{figMOT}) with $P\approx 10^{-6}~$Pascal, the $\kappa\sim 10 ~{\rm sec}^{-1}$ at moderate $R$ limits the number of atoms in the MOT and correspondingly the two-body loss. We therefore expect 
\begin{equation}
N(t)=N_0(1-e^{-\kappa t}).\label{eq:N}
\end{equation}

As being explained in Sec.~\ref{sec:loading}, for small enough $N(t)$ initially, the captured atoms tend to fill the equilibrium locations where radiation pressure is balanced for the nearly empty MOT. The size $\sigma_0$ for this temperature-limited MOT~\cite{Townsend1995} is decided by the MOT restoring force and the equilibrium MOT temperature. Then, as $N(t)$ increases, the atomic sample becomes optically dense to substantially attenuate the MOT beams. This collective modification of radiation pressure is combined with other multi-scattering effects to limit the atomic density at the MOT center, by expelling additional atoms outward. Consequently, the density-limited MOT~\cite{Townsend1995} accommodates more atoms by expanding its size $\sigma(t)$ with $N(t)$. For a moderate-sized MOT as in this work, we expect one-to-one mapping of $\sigma(t)$ and $N(t)$. That is, there is certain functional form 
\begin{equation}
\sigma(t)=\tilde \sigma(N(t),\rho_0),\label{eq:sigma}
\end{equation}
to decide the MOT size according to the atom number. The limiting density $\rho_0$ is largely decided by the intensity and detuning of the MOT beams. 

We expect the Eqs.~\eqref{eq:N}~\eqref{eq:sigma} analysis to be valid for both 2D and 3D MOTs. Here, we assume the 2D-MOT size is isotropic in the $x-y$ plane, with $\sigma$ according to Eq.~\eqref{eq:sigma} and is $z-$independent. Furthermore, We assume the 2D-MOT length, $l=4~$mm, is time-independent. This is supported by the camera video (Fig.~\ref{figMOT}c) during the MOT loading. Physically, the stationary length is associated with the fact that the restoring force along $z$ in the (quasi)2D-MOT is too weak to support any interplay with the multi-scattering force. 

The 2D-MOT is closely aligned to the zero-field line and ONF. But the center is corrugated, due to the MOT beam imbalance~\cite{vengalattore_enhancement_2004,Zhang2012a}. For the 2D-MOT section that is $r_{\perp}$ away from ONF, the atom density at ONF is
\begin{equation}
n(r_{\perp})=\frac{N}{\pi\sigma^2 l} e^{-(r_{\perp})^2/\sigma^2}. \label{eq:n}
\end{equation}
Therefore, with the $r_{\perp}$ distance following certain $g(r_{\perp})$ distribution, the ONF-coupled MOT fluorescence counting rate scales as 
\begin{equation}
\gamma_{\rm SPCM}\propto \int n(r_{\perp})g(r_{\perp}){\rm d}r_{\perp}. \label{eq:gm}
\end{equation}

To facilitate the numerical fit for Fig.~\ref{figMOT}e, in Eq.~\eqref{eq:sigma} we phenomenologically set $\tilde \sigma(N,\rho_0)={\rm max}(\sigma_0,\sigma_s)$, with $\sigma_s=\sqrt{N/(l\rho_0)}$ 
for the density-limited 2D-MOT. The density limit $\rho_0$ is fixed quite accurately first, by comparing the ONF-coupled resonant optical depth for the same MOT sample (Fig.~\ref{figAbs}) with the theoretical estimation based on optical scattering in the near field~\cite{Qiu2022,Ma2023}, leading to $\rho_0=3\times 10^{10}/{\rm cm}^3$. In this step, we use $\tilde \Gamma=2\pi\times 10$~MHz to account for the observed line broadening at the time (Appendix~\ref{sec:lwApp}). By this step, we assume that after the MOT is fully loaded, $\rho_0$ is approximately reached in the near field of ONF.

We simply assume a uniform $g(r_{\perp})$ between $0\leq r_{\perp}\leq b$. With $b$ as auxiliary parameters, we fit the Fig.~\ref{figMOT}e data with Eq.~\eqref{eq:gm} to retrieve $\kappa=(0.4~{\rm sec})^{-1}$. The fit suggests $\sigma_0\approx 0.6 b$ for $t<0.1~$s, and then gradueally expand to $\sigma(t)\approx b$ within one second. The ONF fluorescence is a local observable. Consequently, the atom number $N_0$ retrieved from the measurement depends on the $b$ parameter we choose in the model. For example, $N_0=2.5\times 10^6$ is reached by setting $b=100~{\bf \mu}$m. 

~

~

\section{Details of linewidth measurements}\label{sec:lwApp}


%




\subsection{Summary of linewidth measurements}

As in Fig.~\ref{figAbs}a, we design a composite measurement sequence to probe the atomic gas through the 2D-MOT-ONF interface. Each measurement takes less than 600~ns (Fig.~\ref{figAbs}b), during which the cooling/repumping/heating beams are all shut off. These beams are then switched back on to ensure quasi-continuous operation of the 2D-MOT. To verify that cooling is effective between the nanosecond probes, our measurement results were repeated by increasing $T_{\rm rep}$ to 8~${\bf \mu}$s and 16~${\bf \mu}$s. 

As outlined in Sec.~\ref{sec:spec}, we use two methods to retrieve the atomic response from the transient absorption $T(\Delta, t)$ probed at detuning $\Delta=\omega-\omega_{eg}$. Here $\omega$ and $\omega_{eg}$ are the laser frequency and the atomic transition frequency, respectively. The atomic transition for Fig.~\ref{figAbs} data is $F=2-F'=3$ hyperfine transition of $^{87}$Rb.  For comparison, in this section, as in Fig.~\ref{figApp1}b, we also discuss the linewidth measurements for the nearby $F=2-F'=2$ transition.

The first method to retrieve the transition linewidth follows the standard frequency scan~\cite{Vetsch2010,Patterson2018,Solano2019c} with repeated measurements. We estimate the steady-state optical depth at each detuning $\Delta$ as ${\rm OD}(\Delta) = -\ln\langle T(\Delta, t)\rangle_t$, using transmission data averaged over a time window at sufficiently late times $t \gg 1/\Gamma$. Specifically, for Fig.~\ref{figAbs}d, we average the data over the interval $t_{\rm start} < t < t_{\rm start} + 5~{\rm ns}$, with $t_{\rm start} = 195~{\rm ns}$, to generate the Fig.~\ref{figAbs}c data. A Lorentzian fit is then used to extract the line center and linewidth.
We have verified during data analysis that the fitted lineshape is insensitive to variations in $t_{\rm start}$ for $t_{\rm start} > 160~\mathrm{ns}$, within the $\delta \tilde \Gamma \sim 2\pi\times 100~\mathrm{kHz}$ fit uncertainty (Fig.~\ref{figAbs}c), suggesting that the steady-state response~\cite{Steck2003} is effectively reached for reliable retrieval of the linewidth. This assumption is also supported by single-atom-based simulations~\cite{Qiu2022,Ma2023}. Physically, after the weak probe is switched on, the non-adiabatic transients decay exponentially with $\tilde \Gamma t$. Accordingly, the systematic error associated with the transient response is exponentially suppressed, allowing accurate retrieval of the steady-state response~\cite{Steck2003}.

The second method~\cite{futurePub} is to fit the linewidth directly from a single-shot $T(\Delta, t)$ curve. This method is based on the observation that as long as the ONF-atom response is linear, then, the time-dependent transmission of the probe, such as the Fig.~\ref{figAbs}d data where $\Delta=0$, is decided by the linear spectral response via Fourier transform $\mathcal{F}$, 
\begin{equation}
\begin{aligned}
E_{\rm out}(t)&=\mathcal{F}(E_{\rm in}(\Delta)h(\Delta)),\\
h(\Delta)&=e^{i\phi(\Delta)-{\rm OD}(\Delta)/2}.
\end{aligned}\label{eq:ld}
\end{equation}
By modeling the complex phase shift $\varphi(\Delta)=\phi(\Delta)+i{\rm OD}(\Delta)/2$~\cite{huang2025} via Beer-Lambert law~\cite{Asenjo-Garcia2017}, $\varphi\propto \rho \alpha l$ and $\alpha\propto 1/(\Delta-i\Gamma/2)$, it is possible to retrieve the atomic information directly by fitting the single-shot transmission data via $T(t)=|E_{\rm out}(t)/E_{\rm in}(t)|^2$.  As detailed in ref.~\cite{futurePub}, to efficiently retrieve the atomic response, the probe envelope $E_{\rm in}(t)$ must be complex enough. Therefore, instead of using a simple square pulse (Fig.~\ref{figAbs}a), we keep the probe on for $t>200~$ns, but with its frequency shifted by 250~MHz (not shown in Fig.~\ref{figAbs}). An interference is induced between the coherent forward emission -- see Fig.~\ref{figAbs}b at $t>200~$ns for the intensity profile of the ``flash''~\cite{Chalony2011, Kwong2014} -- with the frequency-shifted probe. The interference data is complex enough for us to infer the transition center and linewidth within a single shot~\cite{futurePub}. 


We verify that the frequency-scan method and the single-shot method agree. The single-shot method, which utilizes the full $T(t)$ profile, is more data-efficient and is exploited to investigate the origin of the line broadening observed in Fig.~\ref{figAbs}(c,d). In the following, we present evidence that the line broadening is related to the surface quality of the ONF~\cite{Nayak2012} and is likely of magnetic origin.

\subsection{On magnetic broadening of absorption profile}\label{sec:mag}
The absorption line of hyperfine transition can be efficiently broadened by a magnetic field, if it is perpendicular to the quantization axis set by light. For example, for the $F=2-F'=3$ transition of $^{87}$Rb probed by an $x-$polarized light, the transition frequency of the $\sigma^{\pm}$ transitions along $z$ is split by as much as 1.9~MHz per Gauss~\cite{Steck2003}. This splitting of transition frequency efficiently broaden the overall absorption profile. Assuming equal population on the $F=2$ Zeeman sublevels, numerical simulation~\cite{Qiu2022,Ma2023} for the $F=2-F'=3$ transition suggests a 1-Gauss field lead to $\sim$1~MHz line broadening while largely keeping the Lorenzian shape. Larger field does distort the line, but we expect that when the field strength has a distribution, then the distortion can be somewhat smoothed out. The broadening is less efficient for the cases where the light-atom interaction commute with Zeeman shifts, such as the $\pi$ or $\sigma$ transition alone.  

The line-broadening effect by the longitudinal field, as described above, depends on the magnetic $g_F$ factors as well as the distribution of $\sigma^{\pm}$ transition strengths between Zeeman sublevel~\cite{Steck2003}. Our numerical simulation suggests that the broadening is approximately halved for the $F=2-F'=2$ transition. This difference of magnetic response motivates us to investigate the two transitions together, as to be detailed in the following.

\subsection{MOT alignment}
Optimal operation of the 2D-MOT requires balancing the intensities of counter-propagating MOT beams~\cite{Townsend1995}. Here, for the prism-MOT as in Fig.~\ref{figSetup}c and Fig.~\ref{figMOT}d,  we achieve the balance by shifting the position of main beam and walk the angles of the small beams. The balance is verified by the insensitivity of the MOT center, according to the direct and prism-reflected ONF-2D-MOT images (Fig.~\ref{figMOT}(b,c)), following a 5-fold reduction of the MOT beam power. The slimmer 2D-MOT at the lower MOT beam power ($\sim$1~mW/cm$^2$ per beam at the $\Delta/2\pi=-12$~MHz MOT detuning) also helps us to verify that the ONF is aligned along the zero-field line to within 50~mrad, at least.

\subsection{Magnetic field scan}\label{sec:magscan}

\begin{figure}[htbp]
        \centering
        \includegraphics [width=1\linewidth]{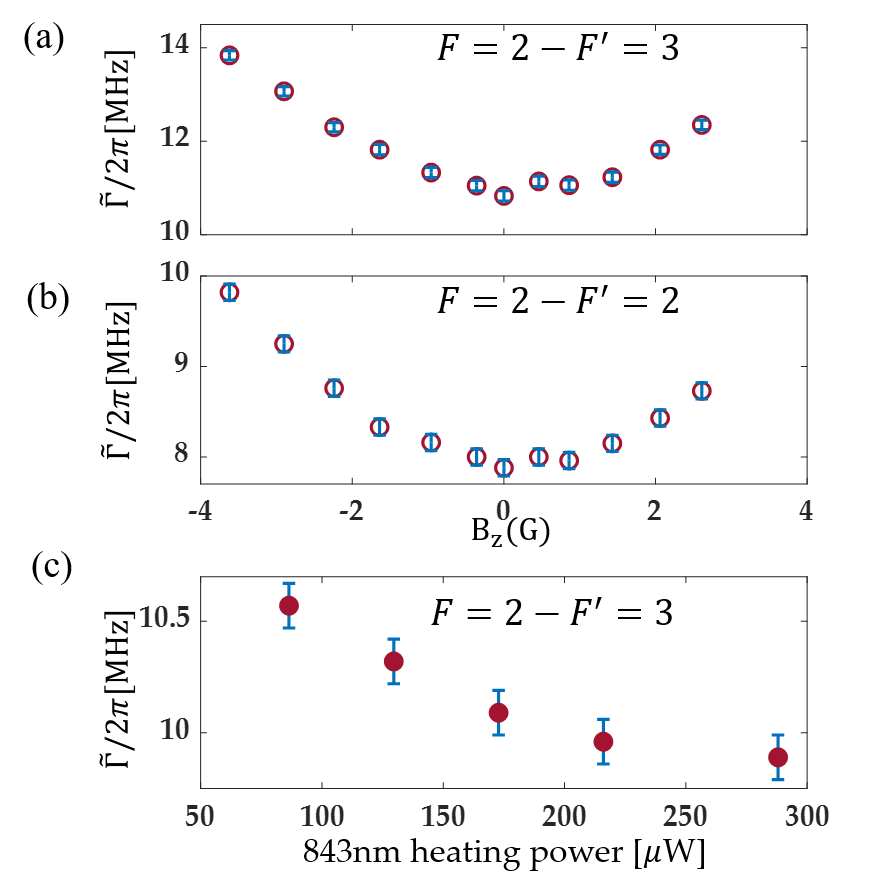}
     \caption{(a) The measured linewidth of $F=2-F'=3$ transition as a function of the $B_z$ offset. (b) The measured linewidth of $F=2-F'=2$ transition as a function of the $B_z$ offset. (c) The evolution of the $F=2-F'=3$ linewidth following an increase of the heating laser power (which is shut off during probe) in a few hours.  }
    \label{figApp1} 
  \end{figure}

We use the plate-coil and $x-$coil as in Fig.~\ref{figMOT}d to shift the $B$-field bias in the $x-y$ plane. We scan the field by up to 1~Gauss in each direction, hardly see a change of linewidth, before the observed absorption signal is too small to support reliable readouts. This is expected. As already suggested by the Fig.~\ref{figMOT}f data, the shift of the MOT center rapidly reduces the number of atoms interacting with ONF in the near field.  We therefore keep the $B_{x,y}$ offsets that optimize the optical depth experienced by the ONF-coupled resonant probe, in all the other measurements. 

We use an additional $z-$coil with diameter $D_z=200~$mm, mounted $\sim$300~mm away, to generate a $B_z$-offset. Similar to the case of $x-$coil, we use Radia~\cite{Elleaume1997,Chubar1998} to confirm that when the $B_z$ offset is generated, the field variation along the $l=4$~mm ONF interaction length is below $1\%$. With the $B_z$ offset, we repeat the single-shot linewidth measurement as by Eq.~\eqref{eq:ld}~\cite{futurePub}. Typical results on the retrieved linewidth, for the case of linearly polarized HE$_{11}$ probe (Appendix~\ref{sec:polarization}), are presented in Fig.~\ref{figApp1}a. For circularly polarized HE$_{11}$ probe, we also find shift of line center at MHz per Gauss level, as expected. Near $B_z=0$, however, the minimal linewidth $\tilde \Gamma/2\pi$ is as large as 10.5~MHz. This linewidth is the smallest we can get at the time of the measurement (See Appendix~\ref{sec:evolve}). We verified that this minimal linewidth is not sensitive to the polarization state of the HE$_{11}$ probe.  


\subsection{Drift of linewidth and heating-induced recovery}~\label{sec:evolve}



The linewidth shown in Fig.~\ref{figApp1}a was measured after the single-chamber ONF-2D-MOT system (Fig.~\ref{figMOT}a) had been in continuous operation for over a year. During the initial six months, we observed a gradual increase in the linewidth, rising from an initial value of $\tilde \Gamma/2\pi \approx 9.2$~MHz to a stabilized minimum of 10.5~MHz, as shown in Fig.~\ref{figApp1}a. To prevent contamination of the ONF from surface adsorptions, a heating laser with $P_{\rm heating} \approx 80~\mu$W at a wavelength of 843~nm was kept on throughout the operation. This laser was always turned off during spectroscopy measurements (Fig.~\ref{figAbs}a) to ensure it did not directly influence the atomic signals.

Interestingly, as in Fig.~\ref{figApp1}c, we observe a reduction in the measured linewidth, following an increase of the 843~nm heating laser power. The measurements were made with one-hour intervals for $P_{\rm heating}$ between 80 and 180~$\mu$W, and 10-minute intervals for the other data points. The heating appears to reset the linewidth back to about 9.7~MHz. This linewidth reduction is not reversible. The $F=2-F'-3$ width stays around 9.5~MHz level for the next weeks, as in Fig.~\ref{figAbs}(c,d), when the heating power is turned down again. The $F=2-F'=2$ linewidth, as below, is accordingly reduced to $\sim$7.5~MHz.  

\subsection{$F=2-F'=2$ transition}

As in Fig.~\ref{figApp1}b, we also perform the $F=2-F'=2$ linewidth measurements during the $B_z$ scan. Comparing to the $\Gamma=2\pi\times 6.1~$MHz linewidth, the amount of broadening for the $F=2-F'=2$ line is roughly half of that for the $F=2-F'=3$ line in Fig.~\ref{figApp1}a. This agrees with the Appendix~\ref{sec:mag} analysis, suggesting that the line broadening is of magnetic origin.  

\subsection{Probe intensity and polarization}\label{sec:polarization}

The optical power of the probe pulse is calibrated by a single-photon detector (Excelitas SPCM-AQRH-16) to be $P_{\rm probe}=5~$pW~\cite{foot:sat}. By doubling the probe power and repeating the measurements for consistency, we verified that the probe is well below saturation in the near field. 

The probe polarization is controlled by a $1/4$ plate and a $1/2$ plate right before the fiber coupling~\cite{Ma2020}. The polarization status of the HE$_{11}$ mode at the ONF section is measured by observing Rayleigh scattering~\cite{Vetsch2012} through direct and mirror-reflected images in our prism-MOT setup (Fig.~\ref{figMOT}b). By scanning the motorized-stage-mounted waveplates and read the Rayleigh scattering along both $x,y$ directions, we are able to model the fiber polarization status quite accurately. This facilitates probe measurements with different HE$_{11}$ polarizations, as in Appendix~\ref{sec:magscan}. 

\subsection{Other factors}
Our probe laser (Moglabs CEL) has a linewidth below 1~MHz. The reliable operation of the OAWG laser probe~\cite{wang_intense_2022} is verified with independent measurements~\cite{huang2025}. The atomic density in the ONF near field is expected to be moderate. By varying the MOT cooling power to reduce the atomic sample size, we verified that the linewidth is consistent for small (OD=0.6) and large (OD=2) optical depth.


\subsection{Discussions}
In this Appendix, we have detailed linewidth measurement in our ferromagnetic 2D-MOT-ONF system. Exploiting the field-free interface, the measurement at $f_{\rm rep}=250~$kHz repetition rate helps us to obtain good signal statistics within minutes. By the fast probe calibration, as presented in Fig.~\ref{figAbs}(a,b), our ONF-spectroscopy is robust against low-frequency technical noises. Exploiting the nanosecond transient probe (also see ref.~\cite{futurePub}), our measurements preserve the atomic state and density distribution in the near field. That we see a broadened atomic line without substantial surface-interaction induced asymmetry~\cite{Sague2007, Nayak2012, Patterson2018, Solano2019c} is quite unexpected.  

The data presented in Fig.~\ref{figApp1} provide key evidence that the line-broadening is not only of magnetic origin but also affected by the heating laser through ONF. Up to the completion of this work, we do not have a definite conclusion on the origin of the line broadening. Nevertheless, from Fig.~\ref{figApp1} we speculate that the broadening could be induced by certain nanoparticles that are magnetic, which are assembled to the ONF surface during prolonged ONF operation in moderate vacuum. Previous study have suggested that atoms adsorbed onto the nanofiber surface may assemble into nano-structures to affect the surface interaction~\cite{Nayak2012}.

\bibliographystyle{revTex2}
\bibliography{ref}

\begin{thebibliography}{67}
\bibitem [{foo({\natexlab{a}})}]{foot:closure}%
  \BibitemOpen
  \bibinfo {note} {In the 4-foil assembly, the $\mu-$metal shielding of source magnetic irregularities is compromised at large working distance $h,h'$. While a nearly ideal 2D qudropole field can stil be formed near the 2D symmetry center, the configuration becomes less robust to source field variations. These variations can be shielded by additional, precisely machined $\mu-$metal enclosures for the sources.}\BibitemShut {Stop}%
\bibitem [{foo({\natexlab{c}})}]{foot:coils}%
  \BibitemOpen
  \bibinfo {note} {Rectangular 2D-MOT coils, if perfectly wound and with ideal current flow, should in principle produce a magnetic field of comparable uniformity. However, the simulated $B_1^{x,y}$ gradients decay more rapidly along the $z$-axis—by about 50\% compared to the $\mu$-metal case. This is because $\mu$-metal tends to become more strongly magnetized near the edges, thereby mitigating the field’s falloff.}\BibitemShut {Stop}%
\end{thebibliography}
\end{document}